\documentclass[prb,twocolumn,showpacs,floatfix]{revtex4}
\usepackage[dvips]{graphicx}

\newcommand{\FeOMo}{\ensuremath{\langle\theta_{\rm Fe-O-Mo}\rangle}}
\newcommand{\FeO}{\ensuremath{\langle d_{\rm Fe-O}\rangle}}
\newcommand{\MoO}{\ensuremath{\langle d_{\rm Mo-O}\rangle}}
\newcommand{\FeMoO}{\ensuremath{\langle d_{\rm (Fe,Mo)-O}\rangle}}
\begin{document}

\title{Effect of band-filling and  structural distortions on the Curie
temperature of Fe-Mo double perovkites}

\author{Carlos  Frontera, Diego Rub\'{\i}, Jos\'e Navarro, Jos\'e Luis
Garc\'{\i}a-Mu\~noz, and Josep Fontcuberta}
\affiliation{Institut  de Ci\`encia de  Materials de  Barcelona, CSIC,
Campus Universitari de Bellaterra, E-08193 Bellaterra, Spain.}
\author{Clemens Ritter}
\affiliation{Institut  Laue Langevin, 6,  rue Jules  Horowitz, F-38042
Grenoble Cedex 9, France.}

\begin{abstract}
By  means  of  high  resolution  neutron  powder  diffraction  at  low
temperature  we  have characterized  the  structural  details of  $\rm
La_{x}Sr_{2-x}FeMoO_6$   ($0\leq   {\rm    x}\leq   0.5$)   and   $\rm
Ca_{x}Sr_{2-x}FeMoO_6$ ($0\leq {\rm  x}\leq 0.6$) series of compounds.
This study reveals a similar  variation of the mean bond-angle \FeOMo\
in both series. In  contrast, the mean bond-distance \FeMoO\ increases
with  La but not  with Ca  substitution.  Both  series also  present a
different evolution of the  Curie temperature ($T_C$), which raises in
the La series and slightly decreases  in the Ca one.  We thus conclude
that the enhancement of $T_C$ in  the La series is due to the electron
filling of the conduction band and a concomitant rising of the density
of states at the Fermi level.

\end{abstract}

\date{\today}


\maketitle


Double perovskites of the type $\rm A_2FeMoO_6$ ($\rm A= Sr$, Ba, Ca),
have been predicted to be half-metallic ferromagnets up to their Curie
temperature  ($T_C$), well above  room temperature.\cite{Kobayashi98a}
This  fact  makes them  very  attractive from  the  point  of view  of
applications  in  magnetoelectronics,  and  have  lead  to  a  growing
interest in  this and  other families of  double perovskites.   In the
ideal double perovskite structure Fe and Mo ions are perfectly ordered
in the B position of the perovskite forming two interpenetrating cubic
sublattices    but,     experimentally,    this    order     is    not
perfect.\cite{cationic_order}  The   low  field  magnetoresistance  is
substantial although it decreases as the temperature approaches $T_C$,
so  in   order  to  enlarge   the  working  range  of   the  potential
applications, the  main goal of  numerous studies has been  to enhance
the  Curie   temperature  of  these  compounds.    In  the  celebrated
manganites, this  objective was achieved by  broadening the conduction
band, that is, by enlarging the Mn-O-Mn bond angle via the appropriate
introduction of large cations in the A-site position of the perovskite
building block.  Using this approach, an enhancement of $T_C$ of about
$100\rm\,K$  has been  obtained when  substituting  Ca by  Sr in  $\rm
La_{2/3}Ca_{1/3-x}Sr_{x}MnO_3$.   Interestingly enough, when  the same
strategy is used in the  double perovskites, a very modest enhancement
of  $T_C$ of  only $20\rm\,K$  has been  obtained.  Indeed,  the Curie
temperatures of  $\rm Ca_2FeMoO_6$ and $\rm Sr_2FeMoO_6$  are of about
$380\rm\,K$\cite{madrid:cfmo}                                       and
$400\rm\,K$\cite{Kobayashi98a,cationic_order} respectively.

\begin{table*}
\caption{Cell parameters  and selected bond distances  and bond angles
of      $\rm       La_{x_{La}}Sr_{2-x_{La}}FeMoO_6$      and      $\rm
Ca_{x_{Ca}}Sr_{2-x_{Ca}}FeMoO_6$ found by  high-resolution NPD data at
$T=10\rm\,K$.  The reported AS  concentration has been obtained by the
refinement of XRPD data at RT.  \FeO\ and \MoO\ have been corrected by
the presence of antisites. Due to the high concentration of AS in $\rm
x_{La}=  0.5$ this  correction is  not realistic  and only  \FeMoO\ is
included.  The agreement  factors of the NPD data  refinement are also
reported.}
\begin{tabular}{lr|c|ccc|ccc|}
  & &  & \multicolumn{3}{c|}{$\rm La_{x_{La}}Sr_{2-x_{La}}FeMoO_6$} &
  \multicolumn{3}{c|}{$\rm Ca_{x_{Ca}}Sr_{2-x_{Ca}}FeMoO_6$} \\
$\rm x_{La}$/$\rm x_{Ca}$ & & 0 & 0.3 & 0.4 & 0.5 & 0.2 & 0.4 & 0.6 \\\hline
SG & &\multicolumn{1}{c|}{$I\,4/m$} & 
\multicolumn{3}{c|}{$P\,2_1/n$} & 
\multicolumn{3}{c|}{$P\,2_1/n$} \\
$a$ (\AA) &  & 5.5549(1) &
5.5903(2)& 5.5916(2)& 5.5887(2) &
5.5639(2)& 5.5453(2)&5.5331(3)\\ 
$b$ (\AA) &  & \multicolumn{1}{c|}{---}&
5.5655(2)& 5.5656(3)& 5.5642(3) &
5.5511(2)& 5.5386(3)&5.5328(3)\\ 
$c$ (\AA) &  & 7.9034(2) & 
7.8622(3)& 7.8684(3) & 7.8680(3) &
7.8484(3)& 7.8336(3) &7.8249(3)\\ 
$\beta$ (deg.) &  & \multicolumn{1}{c|}{---} & 
89.95(2)  &   89.92(2) & 89.90(2) & 
90.00(2)& 90.00(2) &89.99(2) \\
AS (\%) & & \multicolumn{1}{c|}{10} & 
\multicolumn{1}{c}{25}        &        \multicolumn{1}{c}{26}        &
\multicolumn{1}{c|}{40} & 
\multicolumn{1}{c}{5} & \multicolumn{1}{c}{5} & \multicolumn{1}{c|}{6}
\\ 
$\langle d_{\rm Fe-O}\rangle$ & & 2.011(4) & 
2.008(5) & 2.012(5) & \multicolumn{1}{c|}{---} & 
2.006(3) & 1.996(3) & 1.993(3) \\
$\langle d_{\rm Mo-O}\rangle$ & & 1.936(4) & 
1.955(5) & 1.953(5) & \multicolumn{1}{c|}{---} & 
1.945(3) & 1.955(3) & 1.957(3)\\
$\langle d_{\rm (Fe,Mo)-O}\rangle$ & & 1.974(2) & 
1.982(2) & 1.982(2) & 1.984(3) & 
1.975(2) & 1.976(2) & 1.975(2)\\
$\langle \theta_{\rm Fe-O-Mo}\rangle$ & & 172.6(3)& 
167.9(4)& 167.6(4)& 166.3(4) & 
167.6(4)& 165.3(5)& 163.2(6)\\\hline
$\chi^2$(\%) & & 2.4 & 
3.9 & 2.1 & 2.9 &
1.9 & 2.2 & 2.8 \\
$R_B$(\%) & & 4.1 & 
4.4 & 4.6 & 4.7 &
3.4 & 3.6 & 3.9\\ \hline
\end{tabular}
\label{tab-la}
\end{table*}

This streaking dissimilarity  indicates substantial differences in the
nature of the ferromagnetic (FM) coupling in these oxides and suggests
that the  origin of ferro-  \makebox{(or ferri-)} magnetism  in double
perovskites is different from  the double exchange mechanism governing
FM  manganites.\cite{ray:prl}  It  has  been  recently  proposed  that
delocalized  electrons,  antiferromagnetically  coupled  to  localized
magnetic moments  (Fe$^{3+}$: $3d^5$), mediate  the magnetic coupling,
resulting  in a  net FM  interaction between  Fe ions.   In  fact, the
magnetic  properties,  at  high  temperature (above  $T_C$),  of  $\rm
Sr_2FeMoO_6$ can  only be properly described if  the contribution from
delocalized electrons of the conduction band and its antiferromagnetic
(AFM)    interaction   with   localized    spins   are    taken   into
account.\cite{itinerant} A  direct consequence  of this model  is that
the strength of the FM interaction  is governed by (i) the strength of
the AFM coupling  between core spins and itinerant  electrons and (ii)
the density of electrons at  the Fermi level [$D(E_F)$].  This in turn
indicates that a possible way  to modify and eventually enhance $T_C$,
could  be the  filling of  the  conduction band  with doping  electron
carriers.  This approach has been proven to be successful and recently
a substantial  enhancement of $T_C$  by more than $80\rm\,K$  has been
reported      in     $\rm     Sr_{2-x}La_{x}FeMoO_6$      and     $\rm
(Ba_{0.8}Sr_{0.2})_{2-x}La_xFeMoO_6$ series,  where a divalent  Sr ion
is         substituted        by         a         trivalent        La
ion.\cite{Navarro01a,serrate,Navarro03a}  Spectroscopic  photoemission
measurements  have been  used to  show  that indeed  $D(E_F)$ in  $\rm
Sr_{2-x}La_{x}FeMoO_6$      enhances     when      $\rm      x$     is
augmented.\cite{fotoemision} Although these  experiments did provide a
solid confirmation of  the relevant role of the  itinerant carriers in
the  FM coupling  in double  perovskites,  they have  not settled  the
microscopic origin of the observed rising of $D(E_F)$.  The difficulty
arises because in these doped  materials, the La substitution not only
may provide carriers  to the conduction band but  also shall promote a
structural distortion owing to  the different ionic radii of Sr$^{2+}$
and La$^{3+}$  ions.  As a result  of it, the  observed enhancement of
$D(E_F)$ cannot  be exclusively  attributed to a  band-filling effect.
In  sharp contrast  with experiments,  theoretical  analyses, assuming
undistorted lattice, have predicted that $T_C$ may lower upon electron
doping,   in   contrast   with  experiments.\cite{guinea}   Therefore,
discrimination between  these effects  (steric and band-filling)  is a
critical  issue for  understanding  the mechanism  of  FM coupling  in
double  perovskites  and for  the  design  of semi-metallic  materials
having still higher $T_C$.

In order  to solve  this problem  and to elucidate  the origin  of the
$T_C$ enhancement we report here a detailed structural analysis of two
complementary series  of materials: La substituted  and Ca substituted
$\rm Sr_2FeMoO_6$. We show that  this selection of materials allows to
discriminate   between    band-filling   and   structural   distortion
effects. From  the comparison of  the structural and magnetic  data we
conclusively  show  that band-filling  effects  dominate  and are  the
responsible  of  the  enhancement  of  the  Curie  temperature  in  La
substituted compounds.  In  contrast, the structural distortions found
in the Ca case (similar to those found in the La case) reduce $T_C$.


Ceramic samples of  $\rm La_{x_{La}}Sr_{2-x_{La}}FeMoO_6$ ($0\leq {\rm
x_{La}}\leq  0.5$) and  $\rm  Ca_{x_{Ca}}Sr_{2-x_{Ca}}FeMoO_6$ ($0\leq
{\rm  x_{Ca}}\leq 0.6$)  have been  synthesized by  solid  reaction in
adequate   atmosphere.    Stoichiometric   amounts  of   high   purity
($>99.99\%$) $\rm  Fe_2O_3$, $\rm MoO_3$, $\rm  SrCO_3$, $\rm CaCO_3$,
and  $\rm  La_2O_3$ have  been  mixed.   After  the initial  prefiring
treatments to decarbonate the  compounds the powders have been pressed
into  rods.   The final  firing  has been  done  at  $\rm 1250^oC$  in
Ar/H-1\% (followed by slow cooling down). The quality of the compounds
was initially  checked by laboratory X-ray  powder diffraction (XRPD),
using long collecting  times in order to obtain  very good statistics.
The obtained  samples are well  crystallized and single  phased.  Only
small traces  ($\leq 0.8\%$)  of $\rm SrMoO_4$  have been  detected in
some   patterns.   XRPD  data   have  been   used  to   determine  the
concentration of  antisites (AS), defined  as the fraction of  Fe (Mo)
ions  in  the Mo  (Fe)  sublattice (thus  $\rm  AS=  50\%$ means  full
disorder; see Table \ref{tab-la}).

The neutron powder  diffraction (NPD) study has been  done at Institut
Laue Langevin  (Grenoble, France).  High-resolution  NPD patterns have
been collected at  the D2B diffractometer (in its  high flux mode with
$\lambda=  1.594\rm\,\AA$) at $T=10\rm\,K$  using the  standard orange
cryostat. Samples were kept at  this temperature for 15 minutes before
data collection.   Medium-resolution NPD patterns  have been collected
at   D20  ($\lambda=  2.42\rm\,\AA$)   diffractometer  in   the  range
$150{\rm\,K}\leq T\leq 510\rm\,K$.   High-resolution NPD and XRPD data
have   been   analyzed  by   the   Rietveld   method  using   FullProf
program.\cite{JRC93}



In  agreement   with  previous  studies,  we  have   found  that  $\rm
Sr_2FeMoO_6$ NPD pattern can be  very well refined with the tetragonal
$I\,4/m$  space  group (SG).\cite{Diana}  NPD  data  reveals that  the
substitution with  La or  Ca induces a  structural transition.   In La
substituted samples  an orthorhombic splitting  of some peaks  and the
loose  of  the $I$-centering  displayed  by  $\rm Sr_2FeMoO_6$  become
evident.\cite{Navarro01a}  A   change  from  $I\,4/m$   SG  (for  $\rm
Sr_2FeMoO_6$)  to $P\,2_1/n$  SG  for $\rm  x_{La}\geq  0.3$ and  $\rm
x_{Ca}\geq 0.2$  takes place.  In Glazer's  notation, this corresponds
to a  change from  $a^0a^0c^-$ ($I\,4/m$) to  $a^+b^-b^-$ ($P\,2_1/n$)
tilt system.\cite{Woodward97} This can be attributed to the small size
of La$^{3+}$  and Ca$^{2+}$ ions  (when compared to that  of Sr$^{2+}$
ions), which reduces the tolerance factor of the perovskite structure,
thus inducing the  rotation of $\rm FeO_6$ and  $\rm MoO_6$ octahedra.
Most  relevant structural details  obtained by  NPD at  $10\rm\,K$ are
reported in  Table \ref{tab-la}.  The concentration  of AS, determined
from XRPD, is  also given.  In all the cases we  have checked that the
composition is,  within the experimental  error, the nominal  one.  In
particular, no appreciable deviations  from the nominal oxygen content
have been found.   We have also analyzed the  magnetic contribution to
the patterns.   The large  range in $Q$  ($0.4 {\rm  \AA}^{-1}\!\leq Q
\leq 7.8 {\rm  \AA}^{-1}$) of D2B data allows  to resolve magnetic and
structural parameters without correlations between them.  In agreement
with previous  studies, data in  Table \ref{tab-la} indicates  that La
and        Ca        doping       promotes\cite{Navarro01a}        and
removes\cite{Goodenough00a} respectively the degree of Fe/Mo order.


\begin{figure}[b]
\centerline{\includegraphics[width=0.7\columnwidth]{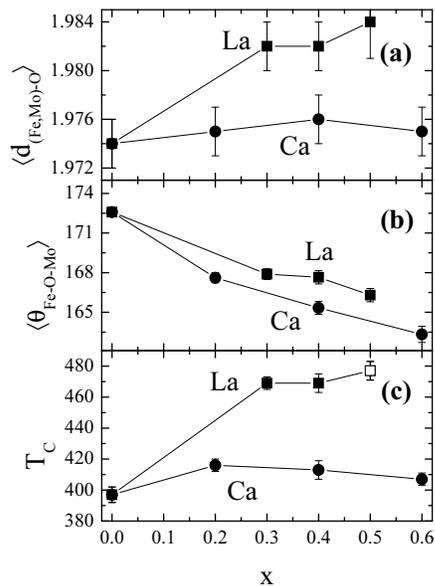}}
\caption{ \FeMoO\ bond-distances {\bf (a)} and \FeOMo\ bond-angle {\bf
(b)}    found    for     $\rm    La_{x}Sr_{2-x}FeMoO_6$    and    $\rm
Ca_{x}Sr_{2-x}FeMoO_6$  at $T=10\rm\,K$.  Solid  symbols in  {\bf (c)}
show  the $\rm  T_C$ found  from data  on Fig.~\ref{TC-NPD};  for $\rm
x_{La}=0.5$ (open square) $T_C$ has been estimated as explained in the
text.  In all panels squares correspond La and circles to Ca series.}
\label{disangtc}
\end{figure}

\begin{figure}[b]
\centerline{\includegraphics[width=0.7\columnwidth]{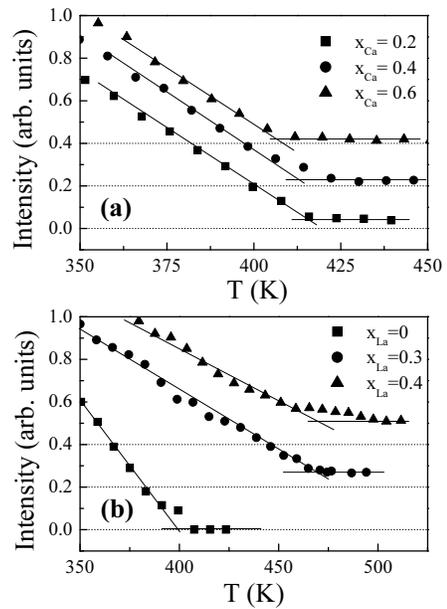}}
\caption{Integrated   intensity  of   $(1\,0\,1)$-$(0\,1\,1)$  doublet
obtained from D20  data for {\bf (a)} Ca and  {\bf (b)} La substituted
compounds. The  straight solid lines below  $T_C$ are a  linear fit to
that  region and  solid lines  above $T_C$  correspond to  the average
value of  the intensity.  Curie temperatures have  been estimated from
the  crossing point  of  the  straight lines.   The  curves have  been
shifted up for clarifying the  picture.  The dotted lines indicate the
zero for,  from bottom to top,  $\rm x_{Ca}= 0.2, 0.4$  and $0.6$ (top
panel), and $\rm x_{La}= 0, 0.3$ and $0.4$ (bottom panel).}
\label{TC-NPD}
\end{figure}

Data  in Table  \ref{tab-la} reveal  a systematic  enlargement  of the
average  \makebox{Mo-O}  bond-distance  (\MoO)  and a  shrink  of  the
average  \makebox{Fe-O}  bond-distance (\FeO)  in  the Ca  substituted
series.  Interestingly enough, the  different sign of these evolutions
almost  compensates  and the  mean  \makebox{(Fe,Mo)-O} bond  distance
(\FeMoO)  remains  nearly  constant  with  $\rm x_{Ca}$.   In  the  La
substituted series  there is  also an enlargement  of \MoO\  but \FeO\
remains nearly constant.  This leads  to an enhancement of the average
octahedral size \FeMoO\ that contrasts with the Ca case as illustrated
in Fig.~\ref{disangtc}(a).  On the other  hand, in both cases there is
a clear bending of the mean \makebox{Fe-O-Mo} bond angle (\FeOMo) with
substitution  [see  Fig.~\ref{disangtc}(b)].    This  is  due  to  the
rotation  of  MO$_6$ octahedra  caused  by  the  smaller size  of  the
La$^{3+}$ and  Ca$^{2+}$ ions when compared to  Sr$^{2+}$. The smaller
size of  Ca$^{2+}$, when compared  to that of La$^{3+}$,  explains the
stronger bond-bending in Ca series (for the same substitution level).

Figure   \ref{TC-NPD}   shows   the   integrated  intensity   of   the
$(1\,0\,1)$-$(0\,1\,1)$  doublet (measured  at D20)  as a  function of
temperature  for   both  La  and  Ca   substituted  compounds.   These
reflections  are  basically  of  magnetic  origin  (although  a  small
structural contribution exists in both $I\,4/m$ and $P\,2_1/n$ SG) and
thus they can be used to  trace the variation of the Curie temperature
upon  substitution.  $T_C$  is  clearly  indicated by  a  kink in  the
temperature dependence of the integrated intensity.
The  remaining intensity  above $T_C$  in the  La case  can  be mainly
attributed to  the presence  of AS and  the resulting AFM  order above
$T_C$.\cite{Diana} Figure \ref{disangtc}(c) collects the values of the
Curie  temperatures  estimated  from  data in  Fig.~\ref{TC-NPD}.   As
already noticed  in Ref.~\onlinecite{Diana}, the  AFM coupling between
nearest-neighbor Fe-Fe pairs occurring in the presence of AS, produces
an  additional magnetic  contribution  to the  $(1\,0\,1)$-$(0\,1\,1)$
doublet above  $T_C$ that masks the  FM onset in NPD  data; indeed the
$\rm   x_{La}=  0.5$  sample   has  $\rm   AS\approx  40\%$   and  the
determination of $T_C$ by NPD data is not accurate.  An enhancement of
$T_C$ of about $8\rm\,K$ with respect to $\rm x_{La}=0.4$ compound was
reported     from     magnetization     measurements.\cite{Navarro01a}
Consequently,  for  $\rm  x_{La}=0.5$  we have  estimated  $T_C\approx
477(6)\rm\,K$  [open square  in Fig.~\ref{disangtc}(c)].   We  note in
Fig.~\ref{disangtc}(c) that Ca  substitution promotes a very moderated
variation of $T_C$:  It rises slightly from $\rm  x=0$ to $\rm x_{Ca}=
0.2$ and lowers gradually for $\rm x_{Ca}>0.2$.  In contrast, there is
an   evident  growth  of   $T_C$  from   $\rm  Sr_2FeMoO_6$   to  $\rm
La_{x_{La}}Sr_{2-x_{La}}FeMoO_6$ ($\rm x_{La}= 0.3$, 0.4 and 0.5).


We  turn  now  to  the  comparison  of  the  variation  of  structural
parameters  and   $T_C$  for  both   series.   In  the  case   of  the
isoelectronically   substituted   Ca-series,    it   is   clear   from
Fig.~\ref{disangtc}(a) and  (b) that with  a monotonic bending  of the
\FeOMo\ bond-angle while keeping the \FeMoO\ bond-distance constant, a
decrease of $T_C$ is  found.  Therefore closing the \FeOMo\ bond-angle
in $\rm A_2FeMoO_6$  leads to a weakening of the  FM coupling and thus
to  a  reduction  of  $T_C$.   The initial  increase  in  $T_C$  ($\rm
x_{Ca}\leq 0.2$) can be attributed  to the reduction of AS observed in
the Ca-substituted samples (see Table \ref{tab-la}).  In contrast, the
La series  shows a radically different behavior.   With similar values
of the  bond-bending, $T_C$ raises  as much as $80\rm\,K$.   Hence, it
follows that  the main difference between the  structural evolution of
both series that can account for  the rising of $T_C$ is the expansion
of the  \FeMoO\ bond in the  La-case.  The enhancement  of the \FeMoO\
bond distance indicates that the substitution of divalent Sr$^{2+}$ by
trivalent  La$^{3+}$  leads  to  the  augmentation of  the  number  of
electrons  within the  metallic sublattice  of the  double perovskite.
Taking into consideration that the available electronic states in both
Fe  and  Mo ions  are  the $t_{2g}$  states  that  participate in  the
conduction band,\cite{Kobayashi98a}  we can asses that  the filling of
this band is effectively enhanced with La doping.

Aforementioned      spectroscopic       measurements      on      $\rm
La_{x}Sr_{2-x}FeMoO_6$  provide evidences that  Mo-band states  at the
Fermi       level       become       gradually       filled       upon
electron-doping.\cite{fotoemision}    Apparently,   data    in   Table
\ref{tab-la},  is  consistent  with  this  finding:  \FeO\  is  nearly
constant   whereas  there   is  a   moderated  enhancement   of  \MoO\
($\Delta\MoO\approx 0.02\rm\,\AA$) upon  La-doping.  In agreement with
Shannon\cite{Shannon76a}  this small  variation  of \MoO\  bond-length
would indicate a predominant electron injection at Mo sites.


It is worth to compare the present behavior with that of well known FM
manganites where $T_C$ is governed  by the bandwidth and by the strong
electron-phonon      coupling     (due     to      the     Jahn-Teller
effect).\cite{Fontcuberta96a,Zhao96,Laukhin97a}  Both  parameters,  in
manganites,  strongly depend  on the  size  of the  A-cations and  the
bending  of  the Mn-O-Mn  bond  angle  drives  a strong  reduction  of
$T_C$.\cite{Fontcuberta96a} As  we have shown here,  the dependence of
$T_C$ on Fe-O-Mo bond angle in double perovskites is much smaller than
in  manganites.  This  fact reflects  that the  strong electron-phonon
coupling present in manganites, is not dominant in the present case.


In conclusion, the main  structural difference between the two studied
series  rely on the  evolution of  the \FeMoO\  bond distance  that is
constant in Ca case but it is enhanced in the La case. The \FeOMo bond
angle  varying similarly  in  both series.   This  indicates that  the
structural distortion caused by La doping does not significantly alter
the FM coupling  in $\rm A_2FeMoO_6$.  The enhancement  of the \FeMoO\
bond distance in La series signals an effective electron doping in the
Fe-Mo sublattice and thus the  filling up of the conducting band.  Our
data  conclusively show  that  the observed  reinforcement  of the  FM
coupling  in  the  $\rm  La_xSr_{2-x}FeMoO_6$ series  originates  from
electron-doping effects rather than from structural ones.
These  findings opens  the possibility  to design  new  strategies for
further enhancement of $T_C$ and shall be of relevance for microscopic
understanding of ferromagnetism in double perovskites.

\begin{acknowledgments}
We thank the AMORE (EU),  MAT 1999-0984-CO3 and MAT 2002-03431 (CICyT,
spanish  government),  and  2001SGR-00334 (Generalitat  de  Catalunya)
projects  for financial  support. C.F.~acknowledges  financial support
from MCyT (Spain).  We thank ILL for the provision beam time.
\end{acknowledgments}

\bibliography{../../tex/bib/cfrontera,%
../../tex/bib/dp,%
../../tex/bib/dp_jose,%
../../tex/bib/gen,%
../../tex/bib/manganites}

\bibliographystyle{apsrev}
\clearpage

\end{document}